\documentclass[twocolumn,tighten]{aastex631}

\newcommand{\halpha}{H$\alpha$}

\newcommand{\hei}{\ion{He}{1}}

\newcommand{\henir}{{\ion{He}{1}}$~\lambda$10830}

\newcommand{\mdot}{$\dot{\text{M}}$}
\newcommand{\Mdot}{{\rm \dot{{M}}}}
\newcommand{\vsini}{$v\sin i$}
\newcommand{\wten}{W$_{10}$}
\newcommand{\msun}{\rm M_{\sun}}
\newcommand{\rsun}{\rm R_{\sun}}
\newcommand{\mstar}{\rm M_{\star}}
\newcommand{\rstar}{\rm R_{\star}}

\newcommand{\msunyr}{\rm{M_{\sun} \, yr^{-1}}}

\newcommand{\kms}{\rm \, km \, s^{-1}}
\newcommand{\ri}{R$_{\rm i}$}
\newcommand{\rw}{W$_{\rm r}$}
\newcommand{\tmax}{T$_{\rm max}$}

\received{May 14, 2022}
\revised{December 14, 2022}
\accepted{December 15, 2022}
\submitjournal{ApJ}
 
\shorttitle{Properties of Low Accretors}
\shortauthors{Thanathibodee et al.}
 
\begin{document}
\title{A Census of the Low Accretors. II: Accretion Properties}
 
\correspondingauthor{Thanawuth Thanathibodee}
\email{thanathi@bu.edu}
 
\author[0000-0003-4507-1710]{Thanawuth Thanathibodee}
\affiliation{Department of Astronomy, University of Michigan, 1085 South University Ave., Ann Arbor, MI 48109, USA}
\affiliation{Institute for Astrophysical Research, Department of Astronomy, Boston University, 725 Commonwealth Ave., Boston, MA 02215, USA}
 
\author[0000-0002-7502-1442]{Brandon Molina}
\affiliation{Department of Climate and Space Sciences and Engineering, University of Michigan, 2455 Hayward St., Ann Arbor, MI 48109, USA}
\affiliation{Department of Atmospheric and Oceanic Sciences, University of Colorado, 311 UCB, Boulder, CO 80309, USA}
 
\author[0000-0001-7351-6540]{Javier Serna}
\affiliation{Instituto de Astronom\'ia, Universidad Nacional Aut\'onoma de M\'exico, Ensenada, M\'exico}
 
\author[0000-0002-3950-5386]{Nuria Calvet}
\affiliation{Department of Astronomy, University of Michigan, 1085 South University Ave., Ann Arbor, MI 48109, USA}
 
\author[0000-0001-9797-5661]{Jes\'us Hern\'andez}
\affiliation{Instituto de Astronom\'ia, Universidad Nacional Aut\'onoma de M\'exico, Ensenada, M\'exico}
 
\author[0000-0002-5943-1222]{James Muzerolle}
\affiliation{Space Telescope Science Institute, 3700 San Martin Dr., Baltimore, MD 21218, USA}
 
\author[0000-0002-1650-3740]{Ramiro Franco-Hern\'andez}
\affiliation{Instituto de Astronom\'ia y Meteorolog\'ia, Universidad de Guadalajara, Avenida Vallarta No. 2602, Col. Arcos Vallarta, CP 44130, Guadalajara, Jalisco, Mexico}

\begin{abstract}
Much is known about the processes driving accretion from protoplanetary disks onto low-mass pre-main-sequence stars (T Tauri stars). Nevertheless, it is unclear how accretion stops. To determine the accretion properties and their relation to stellar properties and gain insight into the last stages of accretion, we present a detailed analysis of 24 low and possible accretors, previously identified using the He\,I\,$\lambda$10830 line. We model moderate-resolution H$\alpha$ profiles of these stars using magnetospheric accretion flow models that account for the chromospheric contribution at the line center. Based on parameters derived from the fits of 20 stars that can be reproduced with the models, we find a power-law relation between the disk truncation radius and the mass accretion rate consistent with predictions from theory and simulations. Comparing the corotation and truncation radii, we find that most of our targets are accreting in the unstable regime and rule out the propeller as the main process stopping accretion. For the truncation radius to be the same as the magnetic radius, the dipole magnetic field and/or the efficiency parameter $\xi$ need to be smaller than previously determined, suggesting that higher-order fields dominate in low accretion rates. Lastly, we determine that the lowest accretion rates that can be detected by {\halpha} line modeling are $1-3\times10^{-11}\,M_{\odot}yr^{-1}$ for M3 stars and $3-5\times10^{-11}\,M_{\odot}yr^{-1}$ for K5 stars. These limits are lower than the observed accretion rates in our sample, suggesting that we have reached a physical lower limit. This limit, $\dot{M}\sim10^{-10}\,M_{\odot}yr^{-1}$, is consistent with EUV-dominated photoevaporation.
\end{abstract}
 
 
\defcitealias{thanathibodee2022}{Paper~I}
 
\section{Introduction} \label{sec:intro}
Detailed modeling of emission line profiles of accreting T Tauri stars (TTS) provides valuable information on accretion properties such as mass accretion rate and accretion geometry \citep[e.g.,][]{muzerolle2001,lima2010,kurosawa2011,alencar2012,thanathibodee2019a,thanathibodee2020}. Comparing the stellar properties with the accretion properties of stars and disks provides insight into the star and disk connection, as well as their evolution \citep[e.g.,][]{manara2016,manara2017a,alcala2017}.
 
In an effort to understand the last stages of the evolution of stellar accretion, we have conducted a survey to search for T Tauri stars near the end of their accretion phase -- the low accretors. These stars still host primordial dust disks at different stages of evolution, but their signatures of mass accretion onto the stars are very weak, indicating that the mass accretion onto these stars has nearly stopped. Using a sensitive probe of accretion -- the {\henir} line -- we identified 51 stars accreting at very low but unknown rates (\citealt{thanathibodee2022}; hereafter Paper~I). We found correlations between accretion properties derived from the {\hei} line and properties of the disks, and in particular, we found that accreting stars, albeit at very low rates, still have dust in the inner disk. We also found evidence that the processes stopping accretion may not be directly related to the stellar mass, suggesting that common processes ending accretion may occur in a wide range of masses. 
 
Although qualitative comparisons between the {\henir} accretion signature and properties of the disks and the stars have provided us with some insight into the last stages of accretion, a more quantitative analysis is needed to get the full picture. In particular, the {\hei} line, although sensitive to accretion, does not provide the measurement of the mass accretion rates or the geometry of accretion. However, this information can be obtained by detailed modeling of emission lines formed in the accretion flows using the magnetospheric accretion model. 
 
Here we report observations and magnetospheric accretion flows modeling of 22 newly identified low accretors and 2 possible accretors to search for correlations between accretion properties and properties of stars and the disks. By quantitatively analyzing the mass accretion rates and geometry, we aim to gain some insight into the last stages of stellar accretion.
 
This paper is organized as follows. In \S~\ref{sec:obs} we describe the observations and data sources. In \S~\ref{sec:results} we derive stellar properties, apply magnetospheric models to fit {\halpha} profiles, and derive the mass accretion ratse of the stars and their geometry. In \S~\ref{sec:discussion}, we discuss the implications of our results. Finally, in \S~\ref{sec:summary} we give our conclusions.

\begin{deluxetable*}{llcccccc}[t]
\tablecaption{Adopted and Derived Stellar Parameters of the Targets}
\label{tab:properties_acc}
\tablehead{
\colhead{Target} 	&
\colhead{Loc} 	&
\colhead{SpT} 	&
\colhead{M$_{\star}$} &
\colhead{R$_{\star}$} &
\colhead{\hei \tablenotemark{a}} &
\colhead{$v\sin i$} &
\colhead{Ref.\tablenotemark{b}} \\
\colhead{} 		&
\colhead{} 		&
\colhead{} 		&
\colhead{(M$_{\odot}$)} 	&
\colhead{(R$_{\odot}$)} &
\colhead{$\lambda10830$} 		&
\colhead{($\kms$)} &
\colhead{}
}
\startdata
2MASS J08074647-4711495   & $\gamma$Vel  & M4    &   0.16    & 0.84  &  b   & $12.9\pm1.6$   & (1) \\ 
2MASS J08075546-4707460   & $\gamma$Vel  & K3    &   1.15    & 1.50  & b,br & $20.5\pm4.1$   & (1) \\ 
ISO-ChaI~52               & Cha~I        & M4    &   0.18    & 1.17  &  r   & $10.46\pm1.31$    & (2) \\ 
CHSM~13620                & Cha~I        & M2    &   0.34    & 1.72  &  r   & $18.9\pm3.5$   & (1) \\ 
SO682                     & $\sigma$Ori  & M0.5  &   0.46    & 1.83  &  r   & $16.58\pm0.79$  & (2) \\ 
CVSO~40                   & OriOB1a      & M0    &   0.53    & 1.47  &  r   & $15.23\pm0.84$  & (2) \\ 
CVSO~156                  & OriOB1b      & M2    &   0.34    & 1.96  &  r   & $8.71\pm1.31$  & (2) \\ 
CVSO~298                  & OriOB1a      & M0    &   0.56    & 1.33  &  br  & $14.56\pm0.49$ & (2) \\ 
CVSO~1295                 & Ori Cloud A  & K7.5  &   0.59    & 1.54  &  c,r & $17.43\pm1.65$  & (2) \\  
CVSO~1545                 & OriOB1b      & M4    &   0.17    & 1.58  &  br  & $9.08\pm1.35$  & (2) \\  
CVSO~1575                 & Ori Cloud A  & K7.5  &   0.57    & 1.65  &  c,r & $12.60\pm1.02$   & (2) \\ 
CVSO~1600E                & OriOB1b      & M3    &   0.27    & 1.65  &  r   & $15.65\pm0.54$ & (2) \\
CVSO~1600W                & OriOB1b      & M3    &   0.27    & 1.54  &  r   & $7.45\pm3.52$  & (2) \\
CVSO~1711                 & Ori Cloud A  & M0    &   0.52    & 1.57  &  r   & $7.18\pm2.77$  & (2) \\ 
CVSO~1739                 & OriOB1b      & M5    &   0.1     & 1.21  &  r   & $14.65\pm1.62$        & (2) \\
CVSO~1763                 & Ori Cloud A  & K3.5  &   1.09    & 1.49  &  r   & $9.21\pm2.05$        & (2) \\
CVSO~1772\tablenotemark{c}& OriOB1b      & M3    &   0.29    & 1.04  &  e   & $11.79\pm1.19$        & (2) \\
CVSO~1842\tablenotemark{c}& OriOB1b      & M2.5  &   0.32    & 1.03  &  e   & $11.72\pm2.13$        & (2) \\
CVSO~1886                 & OriOB1b      & M3    &   0.28    & 1.02  &  br  & $17.48\pm1.07$        & (2) \\ 
CVSO~1928                 & Ori Cloud A  & M0.5  &   0.47    & 1.54  &  r   & $14.20\pm0.42$  & (2) \\ 
CVSO~1942                 & Ori Cloud B  & K6    &   0.72    & 1.57  &  r   & $7.99\pm2.55$  & (2) \\ 
2MASS J16020757-2257467   & USco         & M2.5  &   0.32    & 1.03  &  r   & 14.35        & (3) \\
2MASS J16042165-2130284   & USco         & K2    &   1.39    & 1.98  &  br  & $17.30\pm0.40$        & (4) \\
2MASS J16253849-2613540   & USco         & K7    &   0.56    & 2.46  &  r   & 17        & (5) \\
\enddata
\tablenotetext{a}{Type of {\henir} line profile from \citetalias{thanathibodee2022}: $r$=redshifted absorption, $b$=blueshifted absorption, $br$=blue+redshifted absorption, $c$=central absorption, and $e$=emission.}
\tablenotetext{b}{References for $v\sin i$: (1) \citet{frasca2015}; 
(2) This study, see \S~\ref{ssec:vsini_acc} for details.; (3) \citealt{jonsson2020}; (4) \citet{dahm2012}; (5) \citet{james2006}.}
\tablenotetext{c}{Possible accretors according to the {\henir} lines.}
\end{deluxetable*}

\section{Targets, Observations, and Data Sources} \label{sec:obs}
In this paper, we work with a subset of the newly identified accretors and possible accretors from \citetalias{thanathibodee2022}. In general, we include most stars in the Orion OB1 association and stars from other star-forming regions with available archival spectra. Our targets include stars from Chamaeleon I, $\gamma$~Velorum, $\sigma$Ori, Orion OB1, and Upper Scorpius, and they cover a wide range of stellar parameters. We also include two possible accretors, CVSO~1772 and CVSO~1842, in which the {\henir} lines are in emission without a detectable absorption. Two of our targets, CVSO~1295 and CVSO~1575, are episodic accretors, as their profiles show definite accretion signatures only at some epochs.
 
\subsection{Stellar Parameters}
In Table~\ref{tab:properties_acc}, we list the stars included in our survey, their stellar parameters, the type of their {\henir} profiles, and their rotational properties. We adopted the spectral type, mass, and radius from \citetalias{thanathibodee2022}. We also gathered the projected rotational velocities $v\sin i$ from the literature. For some of the stars in which these data are not available, we determined the {\vsini} from their spectra (c.f., \S~\ref{ssec:vsini_acc}).

\subsection{Optical Spectroscopy}
We obtained optical spectra of 19 stars out of the 51 new accretors identified in \citetalias{thanathibodee2022}, using the MIKE spectrograph on the Magellan Clay telescope at the Las Campanas Observatory in Chile. We use the 0.7'' slit and the red camera, resulting in the spectral resolution of R$\sim$32500 for a wavelength range 4900-9500\,\AA. We reduced the data using the CarPy package \citep{kelson2000,kelson2003}.
 
In addition to the MIKE data, many of the stars in the southern star-forming regions had been observed using instruments on the ESO/VLT. We downloaded the spectra of these stars from the ESO Archive Science Portal. These data had been reduced and calibrated by their instruments' pipeline. Table~\ref{tab:obs_acc} shows the details of the observations and data sources.
 
While the spectral resolutions in these observations are different ($\sim7-16\kms$), they are sufficient to resolve the {\halpha} line; the width of which is of the order of free-fall velocities ($\sim$few hundreds $\kms$). Therefore, we do not expect any significant differences in the line modeling due to the spectral resolution.
 
\begin{deluxetable*}{lcccccc}[t]
\tablecaption{Summary of Observations and Data Sources \label{tab:obs_acc}}
\tablehead{
\colhead{Target} 	&
\colhead{Instrument} 	&
\colhead{Spectral} &
\colhead{Obs. Date} &
\colhead{Exp. time} &
\colhead{SNR\tablenotemark{a}} &
\colhead{ESO Program} \\
\colhead{} 		&
\colhead{} 		&
\colhead{Resolution} 		&
\colhead{(UT)} 	&
\colhead{(sec)} &
\colhead{}      &
\colhead{ID}
}
\startdata
\multicolumn{7}{c}{New Observations} \\ \hline
ISO-ChaI~52 & MIKE      & 32500     & 2020-11-23    & 2311  &  30   & \nodata       \\
SO682       & MIKE      & 32500     & 2020-11-23    &  600  &  30   & \nodata       \\
CVSO~40     & MIKE      & 32500     & 2020-11-23    &  803  &  39   & \nodata       \\
CVSO~156    & MIKE      & 32500     & 2020-11-23    & 1440  &  38   & \nodata       \\
CVSO~298    & MIKE      & 32500     & 2020-11-23    & 1419  &  40   & \nodata       \\
CVSO~1295   & MIKE      & 32500     & 2020-11-23    &  900  &  39   & \nodata       \\
CVSO~1545   & MIKE      & 32500     & 2021-01-17    & 1200  &  24   & \nodata       \\
CVSO~1575   & MIKE      & 32500     & 2020-11-23    &  900  &  39   & \nodata       \\
CVSO~1600E  & MIKE      & 32500     & 2021-01-17    & 1200  &  23   & \nodata       \\
CVSO~1600W  & MIKE      & 32500     & 2021-01-17    &  600  &   9   & \nodata       \\
CVSO~1711   & MIKE      & 32500     & 2021-01-17    &  420  &  26   & \nodata       \\
CVSO~1739   & MIKE      & 32500     & 2021-01-17    & 2250  &  19   & \nodata       \\
CVSO~1763   & MIKE      & 32500     & 2020-11-23    &  630  &  36   & \nodata       \\
CVSO~1772   & MIKE      & 32500     & 2021-01-17    & 1976  &  12   & \nodata       \\
CVSO~1842   & MIKE      & 32500     & 2021-01-17    & 1500  &  16   & \nodata       \\
CVSO~1886   & MIKE      & 32500     & 2021-01-17    & 1440  &  16   & \nodata       \\
CVSO~1928   & MIKE      & 32500     & 2020-11-23    &  720  &  23   & \nodata       \\
CVSO~1942   & MIKE      & 32500     & 2020-11-23    &  835  &  39   & \nodata       \\ \hline
\multicolumn{7}{c}{Archival Data} \\ \hline
2MASS J08074647-4711495\tablenotemark{b}   & GIRAFFE   & 19200     & 2012-01-01    & 6000  &  34   & 188.B-3002(A) \\
   &    &      & 2012-01-02    &   &    &          \\
2MASS J08075546-4707460\tablenotemark{b}   & GIRAFFE   & 19200     & 2012-01-03    & 2400  & 147   & 188.B-3002(A,B)         \\
   &    &      & 2012-02-15    &   &    &          \\
CHSM~13620                & GIRAFFE   & 19200     & 2012-04-30    & 1200  &  67   & 188.B-3002(E) \\
2MASS J16020757-2257467 	& X-shooter 	& 18340 	& 2019-06-07 	& 1400 	&  95 	& 0103.C-0887(B) \\
2MASS J16042165-2130284   & X-shooter  & 18340     & 2019-06-19    &  520  & 187   & 0103.C-0887(B) \\
2MASS J16253849-2613540   & UVES      & 42310     & 2005-04-16    &  780  & 101   & 075.C-0272(A)
\enddata
\tablenotetext{a}{Median Signal-to-noise of the line and the continuum between 6555-6570\,{\AA}.}
\tablenotetext{b}{The spectra for these stars are combined spectra published by the GAIA-ESO survey \citep{gilmore2022}. They were observed on multiple nights, but the line profiles are similar on both epochs.}
\end{deluxetable*}

\subsection{Rotational Velocity} \label{ssec:vsini_acc}
We measured the projected rotational velocities ($v\sin i$) of the stars using the Fourier method, following the procedure of \citet{thanathibodee2020}. We refer the reader to \citet{serna2021} for a more detailed description of the method employed here. In summary, we corrected the spectra by shifting them with the measured apparent radial velocities calculated using the cross-correlation method with template spectra. We then calculated the Fourier power spectra of at least 3 photospheric lines. The number of photospheric lines used for each star depends on the signal-to-noise ratio of its spectrum and the strength of the absorption lines. The $v\sin i$ were then calculated from the location of the first zero in the power spectra. We include the $v\sin i$ measurement in Table~\ref{tab:properties_acc}, along with the stellar parameters.

\begin{deluxetable*}{lcccccccc}[t]
\tablecaption{Range of Model Parameters in the Final Grids \label{tab:model_param_acc}}
\tablehead{
\colhead{Target} & 
\colhead{{\mdot}} & 
\colhead{R$_{\rm i}$} & 
\colhead{W$_{\rm r}$} & 
\colhead{T$_{\rm max}$} & 
\colhead{$i$} & 
\colhead{$a_c$} & 
\colhead{$\sigma_c$} & 
\colhead{no.of.models} \\
\colhead{} & 
\colhead{($10^{-10}\msunyr$)} & 
\colhead{(R$_{\star}$)} & 
\colhead{(R$_{\star}$)} & 
\colhead{($10^4$ K)} & 
\colhead{(deg)} & 
\colhead{} & \colhead{($\kms$)} & \colhead{}
}
\startdata
CVSO~40     & 1.0/2.0/0.1 &  2.6/5.0/0.4 &  0.4/1.6/0.4 &  1.15/1.25/0.025 &  10/80/10 &  0/3.1 &  0/35 & 9240 \\
CVSO~156    & 0.5/2.0/0.1 &  2.0/5.2/0.4 &  0.2/2.0/0.2 &  1.10/1.30/0.025 &  10/80/10 &  0/3.1 &  0/35 & 103680  \\
CVSO~298    & 0.5/2.0/0.1 &  2.0/5.2/0.4 &  0.2/2.0/0.2 &  1.10/1.30/0.025 &  10/80/10 &  0/2.3 &  0/35 & 103680  \\
CVSO~1295   & 0.5/2.0/0.1 &  2.0/5.2/0.4 &  0.2/2.0/0.2 &  1.10/1.30/0.025 &  10/80/10 &  0/1.0 &  0/35 & 103680  \\
CVSO~1545   & 0.5/2.0/0.1 &  2.0/5.2/0.4 &  0.2/2.0/0.2 &  1.10/1.30/0.025 &  10/80/10 &  0/7.3 &  0/35 & 103680  \\
CVSO~1575   & 0.5/2.0/0.1 &  2.0/5.2/0.4 &  0.2/2.0/0.2 &  1.10/1.30/0.025 &  10/80/10 &  0/1.9 &  0/35 & 103680  \\
CVSO~1600E  & 0.8/3.0/0.2 &  2.0/5.2/0.4 &  0.2/2.0/0.2 &  1.10/1.30/0.025 &  10/80/10 &  0/4.2 &  0/35 & 77760  \\
CVSO~1600W  & 0.8/3.0/0.2 &  2.0/5.2/0.4 &  0.2/2.0/0.2 &  1.10/1.30/0.025 &  10/80/10 &  0/16.0 & 0/35 & 77760  \\
CVSO~1711   & 0.5/2.0/0.1 &  2.0/5.2/0.4 &  0.2/2.0/0.2 &  1.10/1.30/0.025 &  10/80/10 &  0/2.5 &  0/35 & 103680  \\
CVSO~1739   & 0.5/2.0/0.1 &  2.0/5.2/0.4 &  0.6/2.2/0.2 &  1.10/1.30/0.025 &  10/80/10 &  0/7.0 &  0/35 & 93312  \\
CVSO~1763   & 0.6/5.0/0.5 &  2.6/5.0/0.4 &  0.8/1.6/0.4 &  1.125/1.25/0.025&  10/80/10 &  0/1.5 &  0/35 & 11592 \\
CVSO~1772   & 0.3/1.5/0.1 &  2.0/5.2/0.4 &  0.2/2.0/0.2 &  1.10/1.30/0.025 &  10/80/10 &  0/2.4 &  0/35 & 84240  \\
CVSO~1842   & 0.5/2.0/0.1 &  2.0/5.2/0.4 &  0.2/2.0/0.2 &  1.10/1.30/0.025 &  10/80/10 &  0/3.4 &  0/35 & 103680  \\
CVSO~1886   & 1.0/4.0/0.2 &  1.2/4.4/0.4 &  0.2/2.0/0.2 &  1.10/1.30/0.025 &  10/80/10 &  0/2.2 &  0/35 & 103680  \\
CVSO~1928   & 0.5/2.0/0.1 &  2.0/5.2/0.4 &  0.2/2.0/0.2 &  1.10/1.30/0.025 &  10/80/10 &  0/1.8 &  0/35 & 103680  \\
CVSO~1942   & 0.5/2.0/0.1 &  2.4/5.6/0.4 &  0.2/2.0/0.2 &  1.10/1.30/0.025 &  10/80/10 &  0/1.7 &  0/35 & 103680  \\
ISO-ChaI~52 & 0.5/2.0/0.1 &  2.0/5.2/0.4 &  0.2/2.0/0.2 &  1.10/1.30/0.025 &  10/80/10 &  0/3.2 &  0/35 & 103680  \\
CHSM~13620  & 0.5/2.0/0.1 &  2.0/5.2/0.4 &  0.2/2.0/0.2 &  1.10/1.30/0.025 &  10/80/10 &  0/1.2 &  0/35 & 103680  \\
J16020757   & 0.3/1.5/0.1 &  2.0/5.2/0.4 &  0.2/2.0/0.2 &  1.10/1.30/0.025 &  10/80/10 &  0/2.2 &  0/35 & 84240  \\
J16253849   & 0.5/1.5/0.1 &  2.6/5.0/0.4 &  0.4/1.6/0.4 &  1.15/1.25/0.025 &  10/80/10 &  0/1.9 &  0/35 & 12320  \\
J08074647   & 0.3/1.5/0.1 &  2.0/5.2/0.4 &  0.2/2.0/0.2 &  1.10/1.30/0.025 &  10/80/10 &  0/2.4 &  0/35 & 84240  \\
J08075546   & 0.5/2.0/0.1 &  2.0/5.2/0.4 &  0.2/2.0/0.2 &  1.10/1.30/0.025 &  10/80/10 &  0/1.3 &  0/35 & 103680  \\
J16042165   & 0.6/3.0/0.2 &  2.0/5.2/0.4 &  0.2/2.0/0.2 &  1.10/1.30/0.025 &  10/80/10 &  0/1.2 &  0/35 & 84240  \\
SO682       & 0.5/1.5/0.1 &  2.6/5.0/0.4 &  0.4/1.2/0.4 &  1.15/1.25/0.025 &  10/80/10 &  0/1.3 &  0/35 & 9240 \\
\hline
\enddata
\end{deluxetable*}

\section{Analysis \& Results} \label{sec:results}

\subsection{Mass Accretion Rates and Accretion Geometry}
After independently determining that the stars are accreting using the {\henir} line, we can be certain that the {\halpha} emission will have some contribution from the accreting material in the magnetosphere.
While it is possible that some stars may be episodic accretors \citepalias{thanathibodee2022} and we did not know if the stars were accreting during the observation, the targets presented here show telltale signs of accretion in the line profiles, e.g., asymmetric, broad wings or redshifted absorption. These features support the identification with accreting stars.

Given that the procedure and precedence for modeling the {\halpha} are well established, and the line modeling is well developed, we used the magnetospheric accretion model to fit the {\halpha} lines of the newly identified low accretors to infer the accretion properties of the stars.
 
\subsubsection{Magnetospheric Accretion Model}
We used the magnetospheric accretion models of \citet{hartmann1994} and \citet{muzerolle1998a,muzerolle2001}, the details of which are described in the original papers. In essence, the model assumes an axisymmetric accretion flow following a dipolar magnetic field. The flow is described by the disk truncation radius {\ri} and the width {\rw} on the disk. The density in the flow is parameterized by the mass accretion rate {\mdot}, whereas the temperature is described by the maximum flow temperature {\tmax}. The line profile calculation is done using the ray-by-ray method for a given inclination $i$.

We created a large grid of models for each star, varying the accretion parameters within ranges appropriate for the observed profile. We started with a large and coarsely spaced grid to search for the approximate location of the reasonably fit model. For a few representative stars, we started the coarse grid with the range of accretion rates from $10^{-10}\,\msunyr$ to $ 10^{-9}\,\msunyr$ and {\tmax} from 10,000\,K to 14,000\,K, and inclinations of {10\degr} to {80\degr}. After identifying that the mass accretion rates for the stars were of the order of $1\times10^{-10}\,\msunyr$ and {\tmax}$\sim$12,000\,K, we focused our parameter space around these values. After defining a new grid and finding the best fits (c.f., \S~\ref{sssec:census2_fit}), we verified that most of the best fits were not at the edge of the parameter space; otherwise, we expanded the parameter space accordingly. Once the approximate parameters were identified, we created the final grid where the parameters are more finely spaced. We adopted the T$_{\rm eff}$=6,000\,K for the accretion shock component on the stellar surface in all cases. This value is less than 8,000\,K used for modeling stars at higher accretion rates, as we expected that there is no veiling from the accretion shock in the continuum since it is closer to the effective temperature of the stars. We verified that removing this shock component (i.e., setting the value to the stellar T$_{\rm eff}$) had no effect on the overall fitting of the models. Table~\ref{tab:model_param_acc} shows the range of parameters used in the final grid of models for each star. From these grids, we determined the best-fit models and inferred accretion parameters from them.

The reader should be mindful of the limitations of the model when applying the results of this work.
In particular, the model assumes no magnetic obliquity and strictly dipolar geometry. Moreover, the temperature distribution in the flow is uncertain. The model that includes magnetic obliquity and non-dipolar fields has not been developed, but simulations with such features exist \citep{romanova2003b,romanova2004a,long2007}. However, running such simulations is computationally much more complicated and beyond the scope of this paper.
 
Some developments have been made to resolve the uncertainty of the temperature distribution in the flow. Using all hydrogen lines observable with the X-shooter spectrograph, Colmenares et al. (in prep.) found that the temperature distribution assumed in our model is consistent with a large number of stars in the Lupus and Chamaeleon I region. Specifically, the hydrogen line decrements \citep[c.f.,][]{antoniucci2017} found in the observations are mostly consistent with that calculated from the same model we used here. Therefore, we argue that the uncertain nature of the temperature distribution would not be a major effect in our analysis.
 
\subsubsection{Fitting the Line Profiles} \label{sssec:census2_fit}
To fit the model to observation, we first removed the photospheric absorption lines from the observed spectrum using a PHOENIX model spectrum \citep{husser2013} with an appropriate effective temperature and gravity for the star. That is
\begin{equation}
    F_{obs} = F_{obs, 0} - F_{phoenix}(T_{eff}, \log g) + 1,
\end{equation}
where $F_{phoenix}$ has already been convolved to the spectral resolution of the observation, and all the spectra involved are normalized to unity.
 
Previous studies show that the chromospheric contribution to the {\halpha} line can be significant in low accretors \citep{manara2017b}, and line profile modeling has to take the chromospheric emission into account. In our previous study \citep{thanathibodee2020}, we approached this problem by fitting the residual between the best fit model profiles and the observation with a Gaussian and found that a Gaussian function can fit the residual qualitatively well. 
This was expected since chromospheric lines, being formed approximately uniformly on the stellar surface, are narrow and symmetric \citep[e.g,][]{batalha1996}. Since we do not attempt to fit the stars' actual chromosphere, a Gaussian function is sufficient for our purpose.  In this paper, we 
account for the chromospheric contribution
by directly incorporating the Gaussian into the fitting procedure. 
 
For each model line flux in the velocity space $F_{v}(\dot{M}, R_i, W_r, T_{max}, i)$, the total model is given by
\begin{equation}
    F_{v, total} = F_{v}(\dot{M}, R_i, W_r, T_{max}, i) + a_ce^{-v^2/(2\sigma_c^2)},
\end{equation}
where $a_c$ and $\sigma_c$ are the amplitude and the width of the Gaussian function representing the chromosphere in the unit of the continuum and $\kms$, respectively. Here we fixed the center of the Gaussian at the stellar rest velocity. We then calculated the fit statistics by varying $a_c$ and $\sigma_c$ to minimize the $\chi^2$,
\begin{equation}
    \chi^2 = \sum_j \frac{\left(F_{j, total}(a_c, \sigma_c) - F_{j, obs}\right)^2}{F_{j, obs}},
\end{equation}
where $j$ indexes over the pixels within $\pm v_{ff,\infty} = \sqrt{2GM_{\star}/R_{\star}}$, the free-fall velocity from infinity, and $F_{obs}$ is the normalized observed flux in the native spectral pixels. 
 
To determine the best fit model in a given grid, we calculated the likelihood,
\begin{equation}
    L = Ne^{-\chi^2/2},
\end{equation}
for each combination of model parameters. $N=e^{\chi^2_0/2}$ is a normalization chosen so that the highest likelihood in the grid is unity and $\chi^2_0$ is the lowest $\chi^2$ in a given grid.
 
\begin{figure*}[t]
\epsscale{1.15}
\plotone{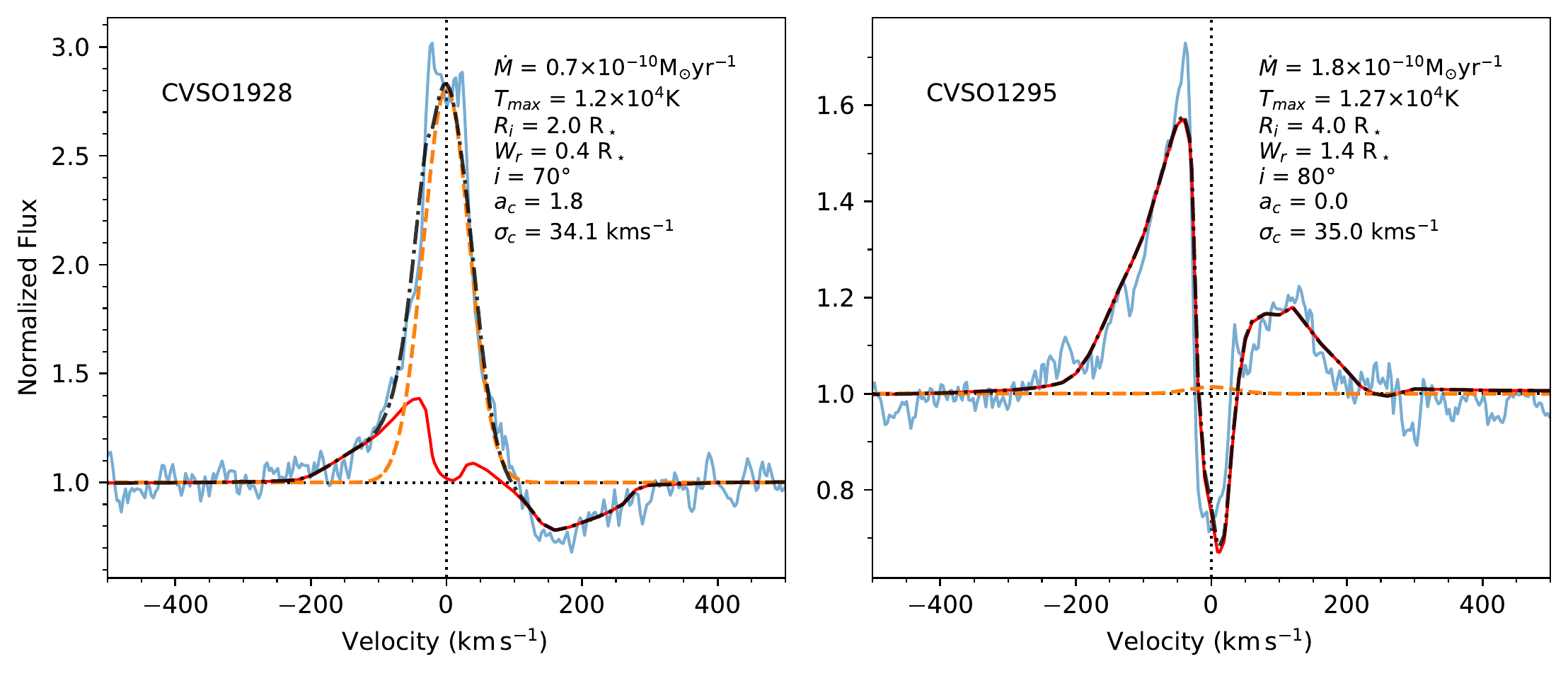}
\caption{Examples of the fitting procedure incorporating the chromospheric emission. The dotted horizontal and vertical lines show the continuum level and the line center, respectively. The solid blue lines are the observation. The solid red lines are the contribution from the magnetospheric model. The dashed orange lines are the best fits for the chromospheric components of the line. The dash-dotted black lines are the best fit total model. 
\label{fig:best_fits_example}}
\end{figure*}
 
Figure~\ref{fig:best_fits_example} shows the examples of the best fits for two stars employing our fitting procedures. The line profile of CVSO~1928, shown on the left panel of the figure, shows a redshifted absorption component and a slightly broader blue wing compared to the red. In this case, the chromosphere contributes significantly to the line profile, accounting for most of the emission at the line center, whereas the magnetosphere contributes mostly to the wing. On the other hand, the magnetospheric contribution can account for most of the line profile of CVSO~1295, shown on the right panel of Figure~\ref{fig:best_fits_example}. The magnetospheric model can simultaneously fit the low-velocity redshifted absorption and wide blue and red wings. Extra redshifted absorption in CVSO~1886 and J16042165 are not accounted for, but, as discussed earlier, their contributions to the overall profiles are small. The accretion flows that cause such features are expected to have much lower accretion rates than the main accretion flows producing the emission profiles \citep{thanathibodee2019a}.
 
Some residual can also be seen near the peak, where the model slightly under-predicts the emission. This could be where the chromospheric emission can contribute. However, in our approach, we assume that the model components are additive, which may not be the case for lines where the magnetospheric flow absorbs light from both the photosphere and the chromosphere. A possible approach to account for such radiative transfer effect is to include a Gaussian component as a proxy for the chromosphere, in addition to the continuum/photosphere from the star, in the profile calculations. However, such an approach is much more computationally expensive than we have adopted here, especially when many stars are involved. Another approach is to fix the chromospheric line \emph{a priori} from independent measurements; this requires knowledge of the formation of chromospheric emission lines. These approaches will be explored in future studies. In any case, the residual is only a few percent of the line peak and should not contribute significantly to the determination of accretion properties.

\subsubsection{Inferring the Accretion Parameters}
 
\begin{figure*}[t]
\epsscale{1.1}
\plotone{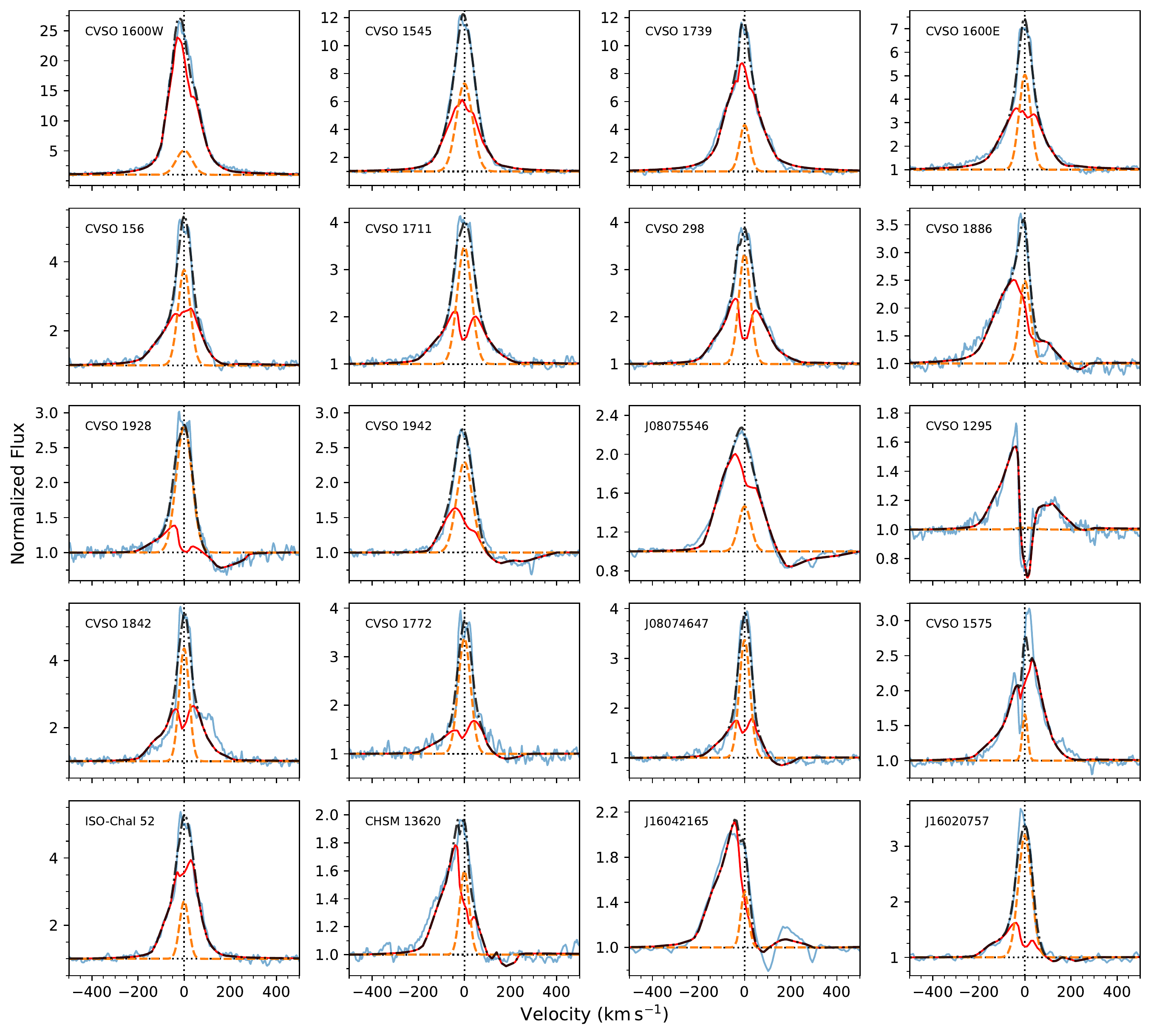}
\caption{The best fits {\halpha} line modeling for stars with good fits. 
Line types and colors as in Figure \ref{fig:best_fits_example}.
\label{fig:halpha_good}}
\end{figure*}

\begin{figure*}[t]
\epsscale{1.15}
\plotone{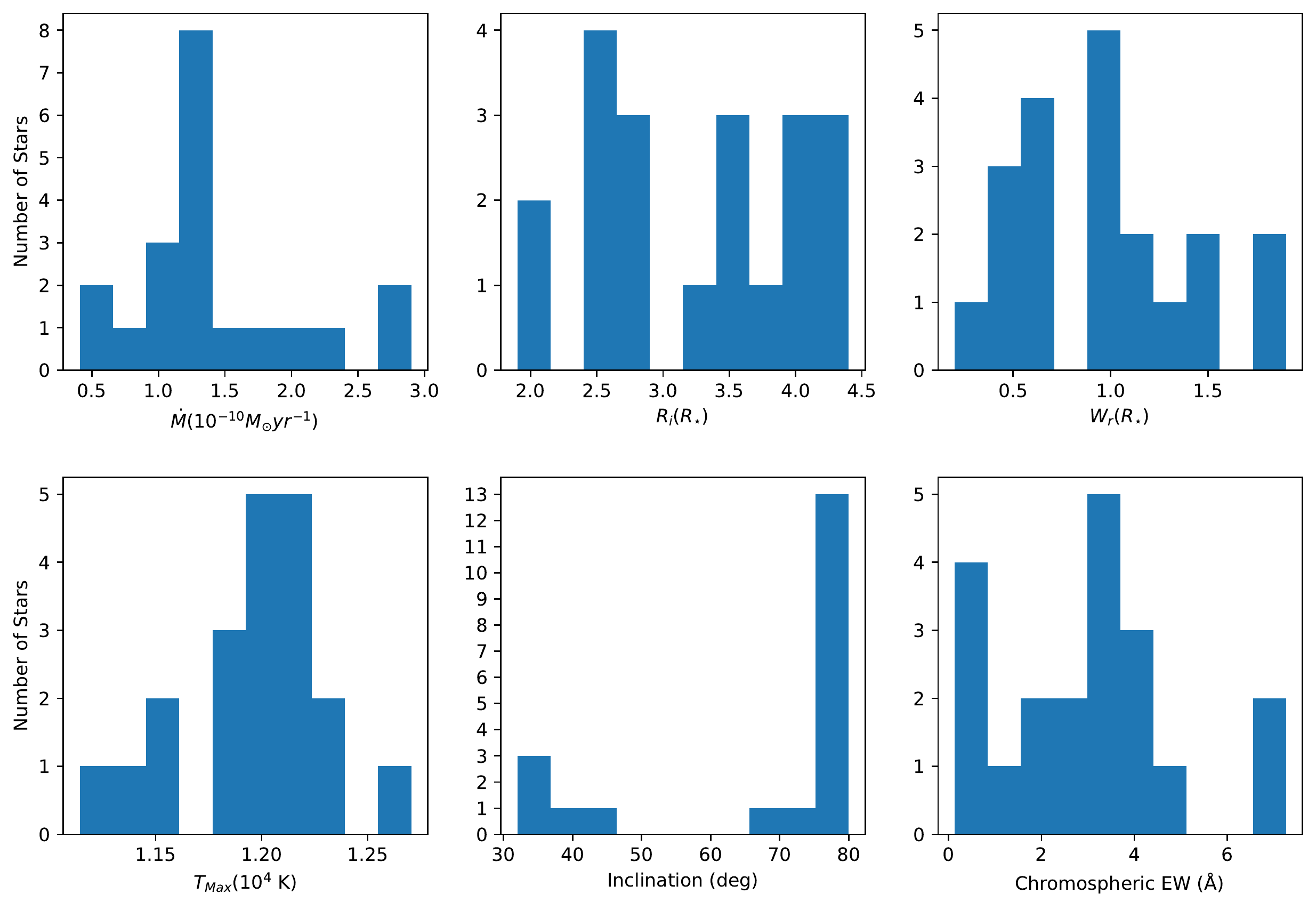}
\caption{The distribution of model parameters for the 20 stars included in the analysis.
\label{fig:parameter_dist}}
\end{figure*}

For each star, we determined the accretion parameters by calculating the weighted mean and standard deviation of each parameter from models with a normalized likelihood of more than 0.5, where we used the likelihood $L$ as the weight. Table~\ref{tab:model_results_acc} shows the accretion parameters retrieved from the magnetospheric accretion modeling and Figure~\ref{fig:halpha_good} shows the {\halpha} profiles where the accretion models can fit the observation qualitatively well. We also list the lowest $\chi^2$ for a given star in Table~\ref{tab:model_results_acc}. The low values of $\chi^2_0$ also confirm that the accretion models fit well. We plot the distribution of the parameters for all the samples selected for analysis in Figure~\ref{fig:parameter_dist}.

\begin{deluxetable*}{lccccccccc}
\tablecaption{Results of the Magnetospheric Accretion Model \label{tab:model_results_acc}}
\tablehead{
\colhead{Star} &
\colhead{\mdot} &
\colhead{\ri} &
\colhead{\rw} &
\colhead{\tmax} &
\colhead{$i$} &
\colhead{$a_c$ \tablenotemark{a}} &
\colhead{$\sigma_c$} &
\colhead{$\chi_0^2$} &
\colhead{Good} \\
\colhead{} &
\colhead{($10^{-10}\,\msunyr$)} &
\colhead{(R$_{\star}$)} &
\colhead{(R$_{\star}$)} &
\colhead{($10^4\,$K)} &
\colhead{(deg)} &
\colhead{} &
\colhead{($\kms$)} &
\colhead{} &
\colhead{}
}
\startdata
CVSO~40\tablenotemark{b} & 2.0$\pm$0.0 & 4.2$\pm$0 & 1.2$\pm$0 & 1.23$\pm$0.0 & 80$\pm$0 & 0.0$\pm$0 & 35.0$\pm$0 & 50.33 & N \\ 
CVSO~156 & 1.4$\pm$0.4 & 2.6$\pm$0.7 & 0.7$\pm$0.3 & 1.2$\pm$0.06 & 79$\pm$3 & 2.7$\pm$0.4 & 25.7$\pm$2.7 & 1.97 & Y \\ 
CVSO~298 & 1.3$\pm$0.5 & 4.3$\pm$0.9 & 0.9$\pm$0.4 & 1.19$\pm$0.06 & 78$\pm$6 & 2.2$\pm$0.3 & 26.9$\pm$3.7 & 0.77 & Y \\ 
CVSO~1295 & 1.5$\pm$0.4 & 4.4$\pm$0.6 & 1.1$\pm$0.4 & 1.22$\pm$0.06 & 79$\pm$0 & 0.1$\pm$0.1 & 24.6$\pm$15.5 & 0.81 & Y \\ 
CVSO~1545 & 1.4$\pm$0.4 & 2.7$\pm$0.9 & 0.9$\pm$0.5 & 1.21$\pm$0.06 & 74$\pm$7 & 4.0$\pm$2.3 & 33.1$\pm$6.5 & 2.21 & Y \\ 
CVSO~1575 & 1.0$\pm$0.2 & 2.0$\pm$0.1 & 0.2$\pm$0.0 & 1.18$\pm$0.06 & 80$\pm$0 & 0.6$\pm$0.2 & 14.6$\pm$9.0 & 3.78 & Y \\ 
CVSO~1600E & 2.3$\pm$0.5 & 2.5$\pm$0.6 & 1.0$\pm$0.4 & 1.18$\pm$0.06 & 80$\pm$0 & 3.5$\pm$0.5 & 25.6$\pm$2.3 & 2.61 & Y \\ 
CVSO~1600W & 2.8$\pm$0.3 & 2.9$\pm$0.2 & 1.5$\pm$0.2 & 1.24$\pm$0.05 & 40$\pm$0 & 4.1$\pm$1.3 & 30.4$\pm$4.9 & 9.52 & Y \\ 
CVSO~1711 & 1.2$\pm$0.5 & 3.5$\pm$1.2 & 0.6$\pm$0.4 & 1.2$\pm$0.06 & 79$\pm$3 & 2.3$\pm$0.3 & 30.7$\pm$3.4 & 0.79 & Y \\ 
CVSO~1739 & 1.7$\pm$0.2 & 2.4$\pm$0.1 & 1.8$\pm$0.2 & 1.22$\pm$0.04 & 80$\pm$0 & 3.1$\pm$0.5 & 21.6$\pm$1.5 & 4.28 & Y \\ 
CVSO~1763 & 3.8$\pm$0.6 & 3.6$\pm$0.5 & 1.2$\pm$0.0 & 1.21$\pm$0.03 & 79$\pm$0 & 0.0$\pm$0.0 & 4.4$\pm$11.6 & 9.84 & N \\ 
CVSO~1772 & 0.4$\pm$0.1 & 4.1$\pm$1.2 & 0.5$\pm$0.2 & 1.14$\pm$0.05 & 80$\pm$0 & 2.4$\pm$0.1 & 24.3$\pm$1.6 & 3.6 & Y \\ 
CVSO~1842 & 1.0$\pm$0.4 & 4.3$\pm$0.9 & 0.9$\pm$0.4 & 1.11$\pm$0.02 & 80$\pm$0 & 3.3$\pm$0.1 & 22.1$\pm$1.2 & 8.32 & Y \\ 
CVSO~1886 & 2.9$\pm$0.8 & 1.9$\pm$0.3 & 1.9$\pm$0.1 & 1.27$\pm$0.03 & 80$\pm$0 & 1.5$\pm$0.1 & 25.5$\pm$2.5 & 2.42 & Y \\ 
CVSO~1928 & 1.3$\pm$0.4 & 3.6$\pm$1.1 & 1.2$\pm$0.5 & 1.2$\pm$0.06 & 32$\pm$17 & 1.5$\pm$0.3 & 32.2$\pm$4.1 & 1.13 & Y \\ 
CVSO~1942 & 1.3$\pm$0.4 & 3.5$\pm$1.0 & 0.7$\pm$0.3 & 1.21$\pm$0.06 & 36$\pm$16 & 1.2$\pm$0.3 & 30.2$\pm$5.4 & 0.88 & Y \\ 
ISO-ChaI~52 & 1.1$\pm$0.4 & 4.1$\pm$0.8 & 1.4$\pm$0.5 & 1.15$\pm$0.05 & 80$\pm$0 & 2.1$\pm$0.4 & 23.0$\pm$2.8 & 1.59 & Y \\ 
CHSM~13620 & 1.4$\pm$0.4 & 2.6$\pm$0.4 & 0.9$\pm$0.3 & 1.22$\pm$0.06 & 79$\pm$1 & 0.7$\pm$0.2 & 22.5$\pm$4.6 & 0.82 & Y \\ 
J16020757 & 0.7$\pm$0.3 & 3.7$\pm$0.9 & 1.3$\pm$0.5 & 1.18$\pm$0.06 & 35$\pm$23 & 2.1$\pm$0.2 & 30.6$\pm$5.1 & 0.36 & Y \\ 
J16253849 & 1.1$\pm$0.2 & 2.7$\pm$0.2 & 0.4$\pm$0.0 & 1.21$\pm$0.05 & 79$\pm$0 & 0.0$\pm$0.0 & 0.2$\pm$0.3 & 136.69 & N \\ 
J08074647 & 0.4$\pm$0.1 & 2.7$\pm$0.6 & 0.7$\pm$0.2 & 1.15$\pm$0.05 & 76$\pm$11 & 2.4$\pm$0.0 & 25.0$\pm$1.8 & 2.41 & Y \\ 
J08075546 & 1.4$\pm$0.4 & 3.2$\pm$0.8 & 0.5$\pm$0.2 & 1.22$\pm$0.06 & 42$\pm$8 & 0.5$\pm$0.1 & 26.7$\pm$8.4 & 0.67 & Y \\ 
J16042165 & 2.1$\pm$0.7 & 4.0$\pm$0.9 & 0.5$\pm$0.2 & 1.23$\pm$0.06 & 66$\pm$5 & 0.5$\pm$0.2 & 22.2$\pm$10.5 & 0.41 & Y \\ 
SO682 & 1.3$\pm$0.2 & 3.6$\pm$0.5 & 0.9$\pm$0.2 & 1.22$\pm$0.03 & 80$\pm$0 & 0.0$\pm$0.0 & 3.7$\pm$5.0 & 13.92 & N \\ 
\enddata
\tablenotetext{a}{$a_c$ is in 
units of the continuum level as the profiles are fitted to the normalized flux.}
\tablenotetext{b}{Too few models to determine uncertainty.}
\end{deluxetable*}
 
\subsubsection{Stars Excluded from Further Analysis}
 
\begin{figure*}[t]
\epsscale{1.15}
\plotone{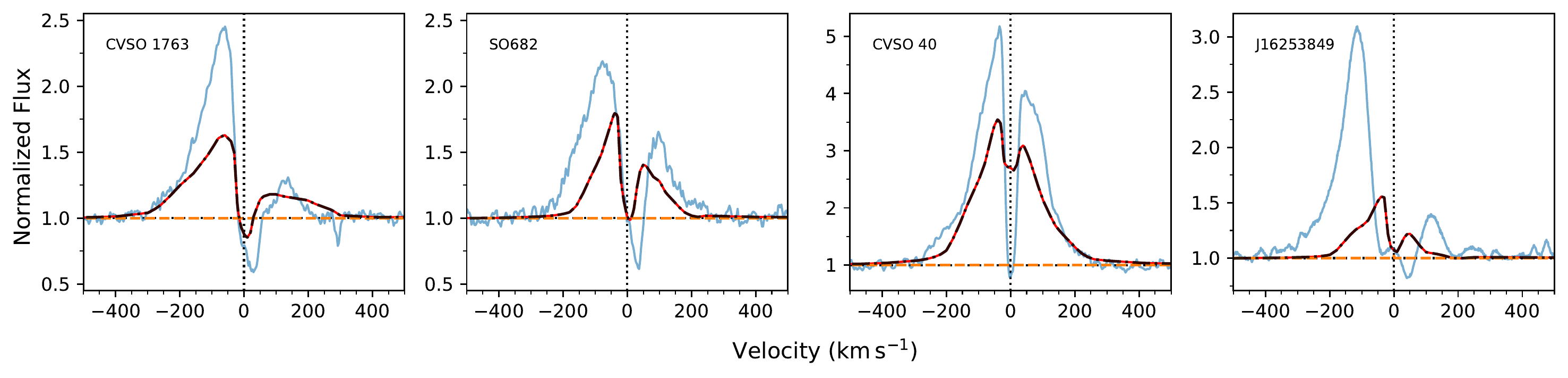}
\caption{Same as Fig.~\ref{fig:halpha_good}, but for stars where the accretion flow model has difficulties reproducing the observations.
\label{fig:halpha_bad}}
\end{figure*}
 
As shown in Figure~\ref{fig:halpha_bad}, the magnetospheric accretion model has difficulty reproducing observations of four stars. All of these stars show low-velocity redshifted absorption features in the line profile, but the magnetospheric model cannot reproduce the absorption feature while producing enough emission to fit the emission component of the line. This is unlike the case of CVSO~1295 (Fig.~\ref{fig:best_fits_example}), in which the emission component is weak, $\sim1.5$ times the continuum, while in these stars the line peaks at $\sim2.5-5$ times the continuum. We speculate that these stars have a flow structure similar to the low accretor CVSO~1335, with its conspicuous low-velocity redshifted absorption in the {\halpha} line in multi-epoch observations \citep{thanathibodee2019a}. Modeling these lines would likely require the two-flow models, as for CVSO~1335. In the interest of comparing the parameters  for large numbers of low accretors, as opposed to performing a detailed study of individual objects, we exclude these four stars from further analysis.

\section{Discussion} \label{sec:discussion}
In this section, we discuss inferences that can be drawn from the accretion rates and flow geometries determined for the sample of low accretors. As explained above, we exclude four stars  that seem to require more complex geometries to fit their {\halpha} profiles  \citep{thanathibodee2019a}. Therefore, the discussion in this section will be based on 20 stars, indicated as `Y' in Table~\ref{tab:model_results_acc}.

\subsection{Accretion Rates and Stellar Properties} \label{ssec:acc_vs_star_prop}
Simulations of accreting magnetized stars have shown that accretion occurs differently depending on the geometry of the magnetic field. Analytical arguments, assuming a dipolar magnetic field, relate the accretion properties and those of the star with the expression
\begin{equation}
    R_M = \xi\left( \frac{\mu^4}{4G\mstar\Mdot^2} \right)^{1/7} \approx 18\xi\frac{B_3^{4/7} R_2^{12/7}}{M_{0.5}^{1/7} \Mdot_{-8}^{2/7}}, \label{eq:rmag}
\end{equation}
where $R_M$ is the magnetic radius, at which the field truncates the disk, $\mu$ is the magnetic moment, $B_3$ is the strength of the surface dipolar magnetic field in kG, $R_2=\rstar/2\rsun$, $M_{0.5}=\mstar/0.5\msun$, $\Mdot_{-8}$ is the mass accretion rate in the unit of $10^{-8}     \,\msunyr$, and $\xi$ encapsulates the details of the disk-magnetosphere interaction \citep{hartmann2016}. From this equation, we expect that the geometry of the accretion flow, represented by the magnetic radius, will depend on the mass accretion rates for similar stellar parameters. 
 
Mass accretion depends on the rotational velocity of the star. One metric used to determine if the system is in the accreting regime or in the propeller regime when accretion does not occur is the fastness parameter, defined as a ratio between the angular velocity of the star and the angular Keplerian velocity at the magnetic radius. It can be written as \citep[e.g.,][]{ghosh1979a,romanova2018}
\begin{equation}
    \omega_s = \frac{\Omega_{\star}}{\Omega_K(r_M)} = \left(\frac{R_M}{R_{co}}\right)^{3/2}, \label{eq:fastness}
\end{equation}
where $\Omega_{\star}$ is the angular rotational velocity, $\Omega_K(R_M)$ is the angular Keplerian velocity at $R_M$, and $R_{co}$ is the corotation radius.
 
In this sub-section, we investigate these relationships to determine the properties of the low accretors.
 
\subsubsection{Accretion Rates and Accretion Geometry}
 
\begin{figure*}[t]
\epsscale{1.0}
\plotone{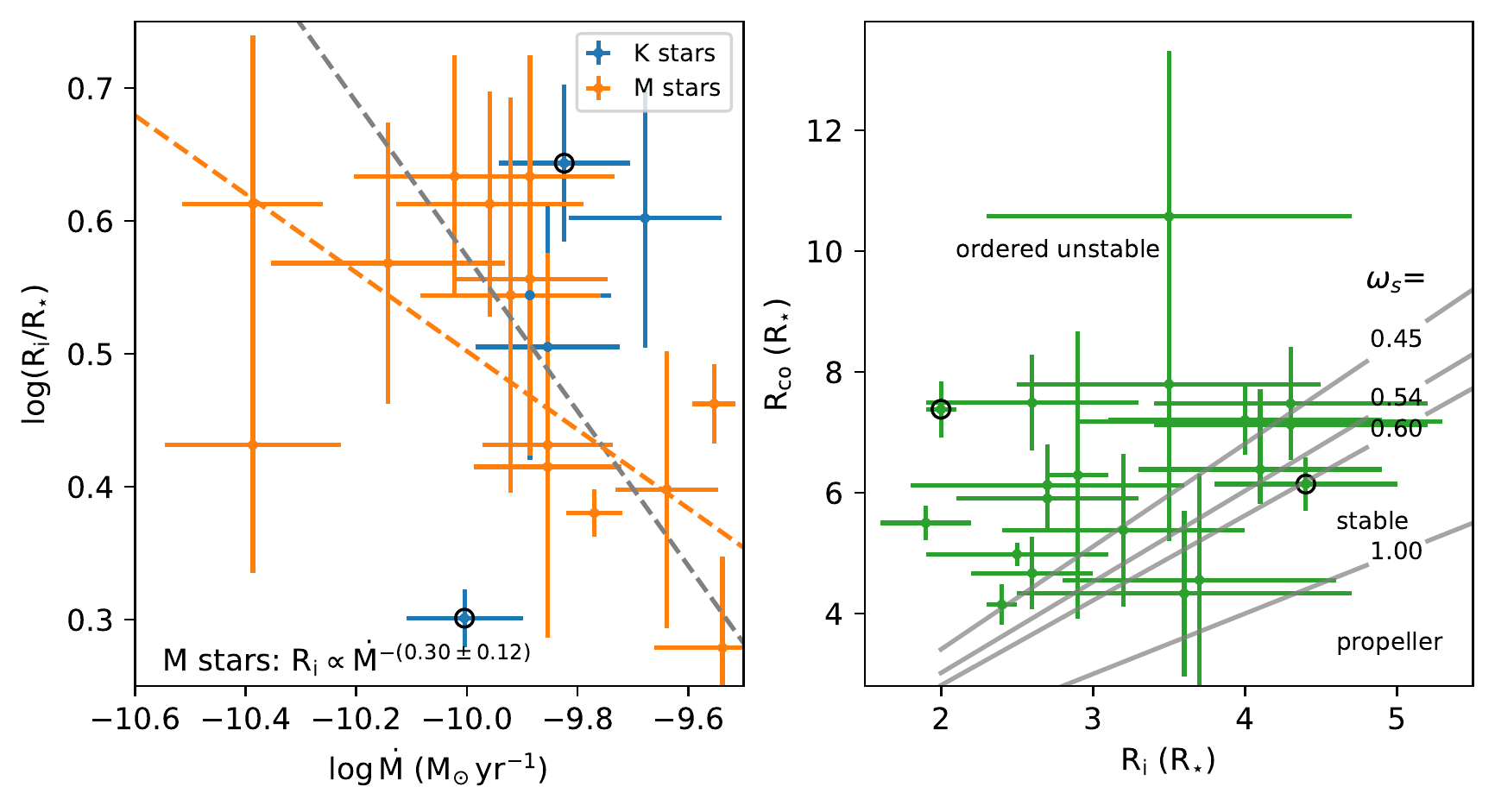}
\caption{Relationships between the truncation radius {\ri} and the mass accretion rate and the corotation radius in low accretors. \emph{Left:} The truncation radius as a function of {\mdot}, plotted in log scale. K-type stars are plotted in blue and M-type stars are plotted in orange. The gray line is the best fit calculated from linear orthogonal distance regression for the whole sample. The orange line is the fit for M-type stars. \emph{Right:} The corotation radius 
as a function of the truncation radius. The gray lines correspond to different radius ratio for different values of fastness parameters $\omega_s$, shown on the plot. The lower right region of the plot is the propeller regime and the upper left region is the ordered unstable regime. The boundary between the stable and unstable regime is $\omega_s\sim0.54-0.6$, depending on the obliquity of the magnetic axis
\citep{romanova2008,romanova2018}.
Stars marked in black circles are episodic accretors.
\label{fig:rin_correlation}}
\end{figure*}

In the left panel of Figure~\ref{fig:rin_correlation}, we plot the truncation radius {\ri} versus the mass accretion rates. Due to the relatively large uncertainty in both parameters, we adopted the orthogonal distance regression \citep[ODR;][]{boggs1987} to determine the trend between the parameters, as opposed to the simple linear regression. This method takes into account the uncertainty in both parameters in calculating the regression and its associated uncertainty. The black dashed line shows the fit for the overall sample, suggesting that the truncation radius increases as the mass accretion rate decreases.

The characteristics of the magnetic fields in T Tauri stars depend on the properties of the stars on their evolutionary tracks \citep[e.g.,][]{villebrun2019}; the magnetic field strength becomes weaker as stars evolve from being fully convective to having radiative cores, for example. This evolution depends on the stellar mass and the age,  which are reflected in the spectral type. In general, earlier spectral type (K-type) T Tauri stars are more massive and/or older than later type T Tauri stars. On the left panel of Figure~\ref{fig:rin_correlation}, we separate the K-type stars from the M-type stars. For the K-type stars, the truncation radii tend to be high, $\sim4\,\rstar$, regardless of the mass accretion rate. Kendall's tau correlation test with p=0.23 suggests that the two parameters are uncorrelated for K stars. However, this result is based on a small sample size and a larger sample is needed to verify it. 
 
On the other hand, for the M-type stars, there is a more significant correlation between the two parameters (Kendall p=0.006) than for the whole sample (Kendall p=0.07). This trend for the M-type stars is expected from the relationship presented in Eq.~\ref{eq:rmag}, suggesting that dipolar magnetic fields still dominate the accretion processes in low-mass low accretors. The same cannot be said for higher mass stars, as no obvious trend emerges. It is possible that as the star gets older (for their masses), the dipolar magnetic fields become weaker, as expected in studies directly measuring magnetic field properties of T Tauri stars \citep[][]{villebrun2019,lavail2017}, making the relationship between {\ri} and {\mdot} less meaningful as it is derived from an assumption of dipolar field.
 
Fitting a linear relationship with ODR to the M-type stars, we found that the mass accretion rates and the truncation radius are related by a power law given by
\begin{equation}
    {\rm R_i} \propto 1/\Mdot^{0.30\pm0.12}.
\end{equation}
This result is robust against the star with the low accretion rates and small {\ri} in our sample (J08074647, log\,$\Mdot =$ -10.40), without which we have the exponent of $0.41\pm0.14$.
 
The exponent $(0.30\pm0.12)$ is consistent with that found by simulations (0.19-0.22, \citealt{blinova2016}; 0.2, \citealt{kulkarni2013}; 0.34, \citealt{ireland2021}), or that derived from equating the disk ram pressure and the magnetic pressure, as in Eq.~\ref{eq:rmag}, $2/7=0.29$. This suggests that in M stars, the assumption of dipolar geometry of the accretion flows is still valid.

\subsubsection{Accretion Geometry and Rotation Properties}
It is known from theoretical studies and simulations that accretion can be in different regimes depending on the relationship between the corotation radius and the disk truncation radius. The fastness parameter $\omega_s$, relating the truncation radius and the corotation radius (Eq.~\ref{eq:fastness}), can be used to classify accreting into different regimes. The system is in the propeller regime if $\omega_s >1$ ({\ri} $>$ R$_{co}$), and steady accretion cannot occur. Instead, some mass will get ejected from the system through winds, while some mass accumulates at the transition region between the two radii and gets periodically loaded onto the star \citep[e.g.,][]{romanova2018}. For non-propellers, accretion may be in a stable or an unstable regime. In the stable regime, accretion occurs in large funnels and the light curve of the star, modulated by accretion hot spots, will be periodic \citep{romanova2008}. Our axisymmetric accretion flow models most resemble these types of flows. The unstable regime, in turn, can be classified into two types based on how the mass is loaded onto the star. In the chaotic unstable regime, mass is loaded in several tongues that could be short-lived, whereas the ordered unstable regime is characterized by tongues merging to form one to two funnels \citep{blinova2016}. 
 
The boundaries between these regimes also depend on the magnetic obliquity, i.e., the misalignment between the rotation axis and the magnetic axis. Accretion is in the ordered unstable regime for $\omega_s\lesssim0.45$ for a wide range of the values of obliquity, and the boundary between the stable and unstable regime is $\omega_s\sim0.54-0.6$ for a misalignment between 5{\degr} and 20{\degr} \citep{blinova2016}.

The corotation radius is typically calculated from the rotation period of the star and the known mass and radius. Since the rotation periods of most of the stars in our sample are not available, we have to infer the period from the projected rotational velocity {\vsini}. Assuming that the inclination of the stellar rotation is the same as the inclination found from the line modeling (Table~\ref{tab:model_results_acc}), we calculated the corotation radius as
\begin{equation}
    R_{co} = \left[  \frac{GM_{\star}R_{\star}^2\sin^2i}{(v\sin i)^2}  \right]^{1/3}. \label{eq:rco}
\end{equation}
Since the assumption of axial symmetry is not necessarily correct, we adopted a minimum uncertainty of 15{\degr} for the inclination unless the modeling results suggest a larger value. The uncertainties in the final values are calculated using standard error propagation methods.
 
The right panel of Figure~\ref{fig:rin_correlation} shows the relationship between the truncation radius and the corotation radius. Four solid lines show the boundary between accretion regimes based on the value of the fastness parameter. None of our targets is in the propeller regime, including the possible accretors (CVSO~1772, CVSO~1842) and episodic accretors (CVSO~1295, CVSO~1575) in our sample. A likely explanation is that these stars are accreting most of the time, but the accretion detection using {\henir} in \citetalias{thanathibodee2022} has some limitations, as discussed therein. 
 
Most of the targets are in the unstable regime, with about half of them in the ordered unstable regime. Given that the gap between the ordered unstable regime and the stable regime is small, it is likely that most of our targets are in the regime in which large funnel flows are present, and applying the magnetospheric model to these cases is appropriate. 
 
These results suggest that low accretors, except for having large truncation radii, are accreting ``normally'' and they are not close to the propeller regime. As noted above, this interpretation is dependent on the assumption that there is no magnetic obliquity. In reality, corotation radii calculated from Eq.~\ref{eq:rco} are likely to be upper limits, as the mean inclination should be $\sim60$\degr. Nevertheless, the $R_{co}$ depends on $\sin^{2/3}(i)$, so the effect of inclination is small at a moderate to high inclination. Therefore, there could be more stars in the stable regime, but it is unlikely that any stars would be in the propeller regime. In addition, the assumption that stars accrete via dipolar fields may influence the interpretation of the results in the accreting regime (however, see Section~\ref{sssec:disk_mag_interact}), but the geometry of the magnetic field should not affect the boundary between accreting and propeller regime.
 
While the exact region on which a particular star is accreting may be uncertain, we can infer that
accretion stops not because the system reaches the propeller regime but from other processes. However, another possibility is that our samples do not include stars nearing the propeller regime because they were not identified as accretors even with the {\henir} line. Testing this hypothesis would require observing stars identified as possible accretors in multi-epoch observations.

\subsubsection{The Disk-Magnetosphere Interaction of Low Accretors} \label{sssec:disk_mag_interact}
Equation~\ref{eq:rmag} provides an explicit relationship between accretion rates and accretion and stellar properties, as discussed above. Given that $R_M=R_i$, the truncation radius, we can use the measurements from our sample to determine the average values of the magnetic field $B$ and the efficiency parameter $\xi$.
 
\begin{figure*}[t]
\epsscale{1.0}
\plotone{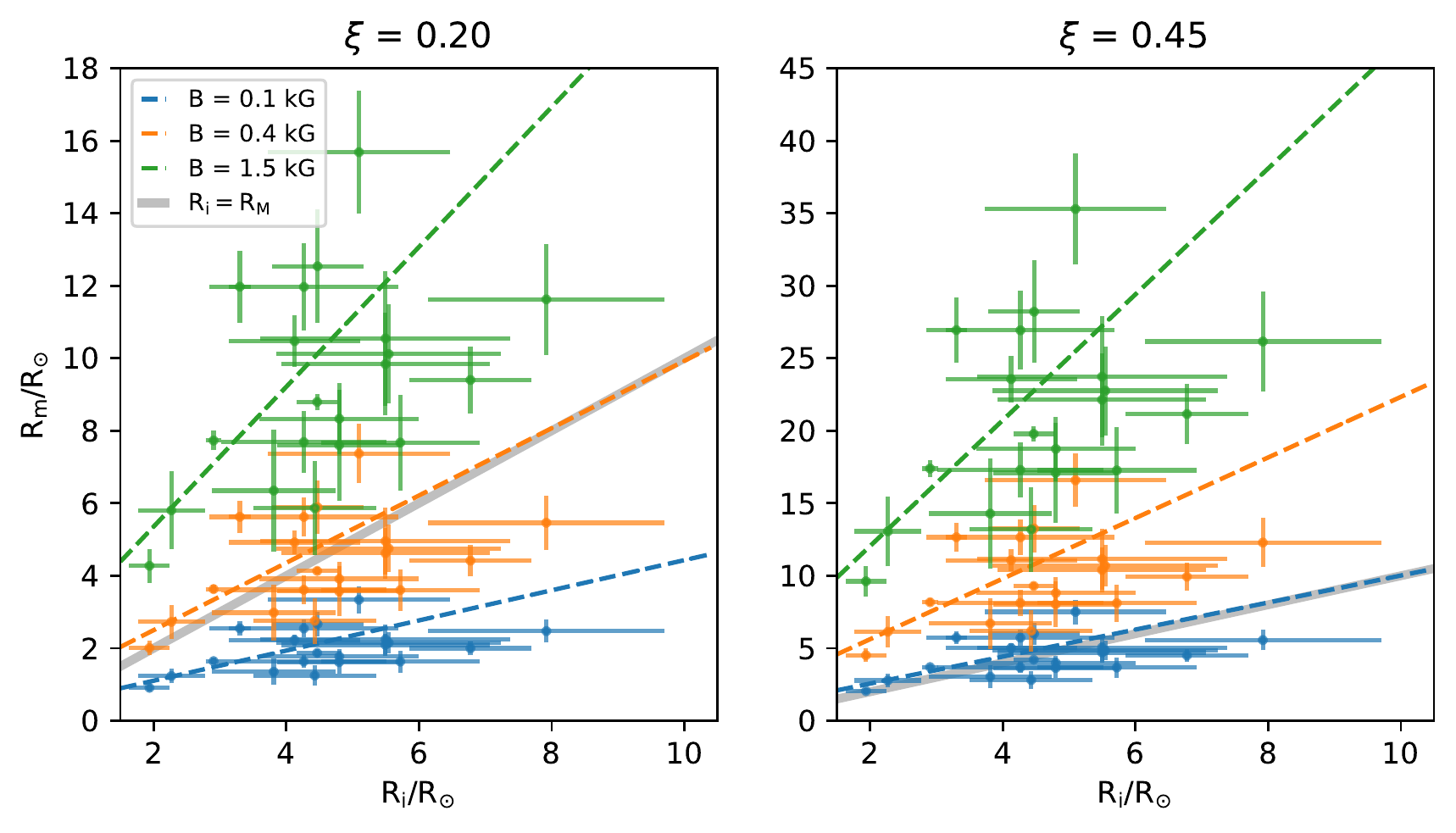}
\caption{The magnetic radii calculated from mass accretion rates and the inferred truncation radii from the model, varying $\xi$ and the surface magnetic field strength B. Different colors refer to different magnetic field strengths. The dashed lines are the linear regressions of the two radii. The solid gray lines are the one-to-one relationship between the two measurements.
\label{fig:Ri_Rm}}
\end{figure*}

Figure~\ref{fig:Ri_Rm} shows the relationship between the truncation radius {\ri} inferred from the magnetospheric accretion flow model fitted to observations and the magnetic radius $R_M$ calculated from Eq.~\ref{eq:rmag} using the inferred {\mdot}. Different colors in the plot refer to different strengths of the surface dipolar magnetic field.
 
The dashed lines are the best linear fits calculated using the orthogonal distance regression, as in Figure~\ref{fig:rin_correlation}, and the solid gray lines are the one-to-one relationship. While {\ri} and $R_M$ show weak correlation, with Kendall p=0.07, it is suggestive, from these fits, that for the magnetic radius to be the same as the truncation radius, the stars generally require a low value of $\xi$ or a low value of the magnetic field B.
 
We can compare these values to that of theoretical expectation ($\xi\sim1$), and numerical simulations. For example, \citet{kulkarni2013} and \citet{blinova2016}, using the large suites of models of their numerical simulations, investigated the relationship between the magnetic radius calculated from the models and that calculated from theoretical prediction such as that presented in Eq.~\ref{eq:rmag}. Their models are for truncation radii between $\sim2-6\,\rstar$, which are appropriate for comparison with our data. \citet{blinova2016} found that the parameter $\xi$ ranges from $0.5\lesssim \xi \lesssim 0.9$ for stable accretion and $0.8\lesssim \xi \lesssim 0.93$ for unstable accretion; a similar range is found by \citet{kulkarni2013}.
 
 For our estimates to be consistent with the high values of $\xi$ obtained in the simulations, the magnetic field strength needs to be as low as 0.1 kG or even lower, as the analysis in Figure \ref{fig:Ri_Rm}  indicates. Given that the total magnetic field strength in T Tauri stars is on the order of a few kG \citep[e.g.,][]{lavail2019}, this suggests that the magnetic field  that is responsible for accreting mass, i.e. the dipolar field, is very weak compared to other topologies (e.g., quadrupole and octupole). Weaker dipole components compared to higher order fields are found in observations \citep[e.g.,][]{gregory2008,donati2011a,chen2013}, but not as much as that of our results. It is possible that as the star evolves, most of the magnetic field lines are in the higher order configuration, but we can still see accretion as if the dipolar field dominated because the mass accretion rate is so low such that even a weak field is able to truncate the disk. To test this hypothesis,  measurements of the components of the magnetic fields in very low accretors would be required.
 
 We also note that the simulations suggesting high values of $\xi$ are run for stars accreting at relatively high accretion rates, $\sim10^{-7}\,\msunyr$. It is  possible that at low accretion rates, the efficiency for the magnetic fields to truncate the disk decrease significantly. Whether the two parameters are directly related, or they result from a common cause (e.g., the evolution of stars/disks) is unclear. Simulations of accreting stars at very low rates are needed to explore this scenario.

\subsection{The Distribution of Accretion Parameters} \label{ssec:model_param_distribution}
 
To search for common properties among the low accretors, we plot the distributions of the best fit model parameters for the 20 stars in Figure~\ref{fig:parameter_dist}. The mass accretion rates and the truncation radii are discussed separately in \S\ref{ssec:acc_vs_star_prop} and \S\ref{ssec:acc_stop}. Here, we discuss the findings and implications for other parameters.

\subsubsection{Inclination}
Perhaps the most surprising finding in these distributions is that the inclinations of low accretors are very high. It is possible that this is in part due to the assumption that there is no magnetic obliquity, and the inclinations of the stellar rotation axes are somewhat smaller than $\sim80\degr$. Tilted magnetospheres should produce light curves that are different from aligned magnetospheres \citep[e.g.,][]{romanova2004a}. Photometric monitoring of these stars will help to test this hypothesis.
 
Another possibility for the high inclinations is that it is an observational bias in the survey to search for low accretors \citepalias{thanathibodee2022}, which used the equivalent width and the width at 10\% peak (\wten) as criteria for selecting low accretor candidates.
 
\begin{figure*}[t]
\epsscale{1.15}
\plotone{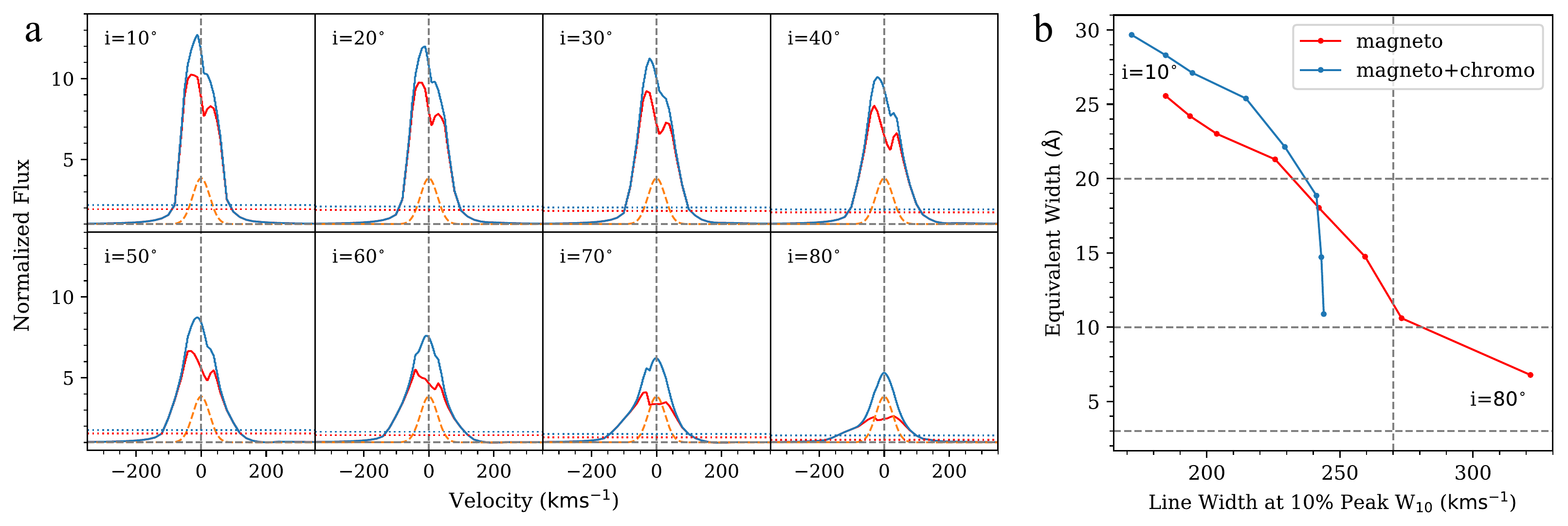}
\caption{Changes in line strength and morphology due to inclination. We adopted the stellar parameters of the M2 star CVSO~156 with model parameters {\ri}=$2.6\,\rstar$, {\rw}=$0.4\,\rstar$, {\mdot}=$9\times10^{-11}\,\msunyr$, and {\tmax}=$1.15\times10^4$\,K. (a) The line profiles at different inclinations from $i=10\degr$ to $i=80\degr$. The red lines are the profiles of the magnetospheric model. The orange lines are the chromospheric contribution, assumed to be constant for all inclinations. The blue lines are the sum of the two components. The dotted horizontal lines show the level at 10\% of the peak flux. (b) The equivalent width and the width at the 10\% flux (\wten) corresponding to the profiles in (a). The same color scheme is used. The points at the upper left of the plot are those with low inclinations, and the points at the lower right are at high inclinations. The horizontal lines are the accretor EW thresholds for spectral type K5, M2.5, and M5.5, and the vertical line is the {\wten} threshold of $270\,\kms$ \citep{white2003}.
\label{fig:prof_incl}}
\end{figure*}

We illustrate this bias in Figure~\ref{fig:prof_incl}, in which we show {\halpha}  profiles calculated for an M2 low accretor observed at different inclinations. The spectral type of the star is the median spectral type of the newly identified accretors in \citetalias{thanathibodee2022}. We assumed that the chromospheric contribution to the line, modeled as a Gaussian, stays constant at EW=$4.1$\,\AA, i.e., more than 50\% lower than \citet{white2003}'s threshold of accretion for an M2 star. As shown by panel (a) of the Figure, the line profiles change both in strength and shape as the viewing inclination changes. In particular, at a low inclination, the total line profile (shown in blue) is dominated by the emission from the magnetosphere. The line is also brighter since few regions of the flow are occulted by the star, the flows in front, or the disk. However, the line is narrow at low inclinations, as the line of sight is closer to normal with the poloidal velocity field of the accretion flow. The line becomes weaker and broader at high inclinations, due to the same effects discussed above. In this limit, the chromosphere dominates the emission line.
 
When observations are made to determine accretion status, no chromospheric contribution is removed, as it is unknown. Therefore, the measurement of EW and {\wten} will intrinsically include the effect of the chromosphere. Panel (b) of Figure~\ref{fig:prof_incl} shows the measurement of EW and {\wten} for the profiles in panel (a) with and without the chromospheric contribution. For reference, the thresholds for accretion classification, EW $\ge \,$10\, {\AA} and  {\wten} $\ge270\,\kms$
\citep{white2003}, are indicated in the Figure.  Without the chromospheric contribution, the star would have been classified as accretor if the EW criterion 
had been used, and if it had an inclination of 45{\degr} or less; at higher inclinations, the EW would be below, or at least barely above, the threshold. On the other hand, it would be classified as an accretor at high inclination if the {\wten} criterion had been used.
 
Taking into account the chromosphere, a similar trend is found for equivalent widths, except that it would need to have a high inclination, $\geq 60${\degr}, to be classified as a non-accretor. For {\wten}, it would always be classified as a non-accretor, regardless of inclination. This is because, at a high inclination, the peak is raised by the chromospheric contribution, so the measurement at the 10\% peak is at the narrower part of the line profile.
 
In \citetalias{thanathibodee2022}, stars selected as low accretor candidates are those classified as non-accretors using either {\wten} or EW. Since most of the targets were selected based on the equivalent width of {\halpha}, the survey preferentially includes stars at higher inclination as those are stars with the highest chance of being (mis)classified as non-accretors. Therefore, it is not surprising that most of our targets show high inclinations in magnetospheric modeling.
 
While the above explanations are applicable to cases with high inclinations, there are still some stars in our sample with low inclinations, but none have moderate inclinations. There are five stars with fitted inclinations of less than 45$\degr$, including CVSO~1600W, CVSO~1928, CVSO~1942, J16020757, and J08075546. As shown in Figure~\ref{fig:halpha_good}, the best fits for most of these stars (except CVSO~1600W) show that the {\halpha} line profiles are dominated by chromospheric emissions with subtle broad wings. 
It could be that at a low inclination, more of the chromosphere could be seen as it is not occulted by the magnetospheric flows. Therefore, with low accretion rates and low inclinations, the signatures of accretion in the {\halpha} line can be hard to detect, and there is a chance that the accretion status can be misclassified compared to accretors observed at moderate inclinations. 
 
Lastly, the derived high inclinations in our analysis could be due to other unaccounted factors in the modeling. To quantify these effects, a comparison between the inclination derived from the modeling and those calculated from rotation periods will be needed. However, the rotation periods of most of the samples are not available, and deriving periods from light curves would be outside the scope of this paper. Nevertheless, we can qualitatively estimate how likely it is for the samples to have high inclinations. For a star with $R=1.5R_{\odot}$ (a typical value for our sample) and a period of 5.84 days (a typical value for WTTS in Orion OB1; \citealt{karim2016}), we get {\vsini}$\sim13\,\kms$ for $i=90\degr$. An accreting system is expected to have a longer period and hence even lower {\vsini}. Therefore, the measure {\vsini} of our samples ($\sim7-20\,\kms$) are not inconsistent with high inclinations found in the modelings.
 
\subsubsection{The Equivalent Width of the Chromospheric Emission}
For the parameters related to the Gaussian chromospheric contribution, we plot instead the equivalent width of the chromospheric line, calculated as 
\begin{equation}
    EW_c = \sqrt{2\pi}a_c\sigma_{c,\lambda},
\end{equation}
where $\sigma_{c,\lambda}$ is the Gaussian width converted to wavelength scale.
 
As shown in the last panel of Figure~\ref{fig:parameter_dist}, most of the stars have the EW of the chromospheric component less than 5\,{\AA}, and a portion of them require very small or almost no chromospheric contribution. We compared the EW of the chromospheric emission to that of the observation and found that, on average, the chromospheric emissions account for $\sim30\%$ of the overall profile. For three targets (CVSO~1928, J16020757, J08074647), the chromospheric emission can be up to $\sim60\%$ of the line. These results confirm previous findings that the chromospheric contribution can be very significant in the line \citep[e.g.,][]{manara2013}.

\subsection{The Lowest Measurable Mass Accretion Rates} \label{ssec:lowest_mdot_model}
The distribution of mass accretion rates of our sample, shown in Figure~\ref{fig:parameter_dist}, suggests that most of the stars accrete at $\sim10^{-10}\,\msunyr$, with the lowest at $\sim5\times10^{-11}\,\msunyr$. Is this a physical limit, or is it the lowest rate that can be measured using line modeling? 
 
To address this question, we ran two grids of models for two 5\,Myr stars, a K5 star and an M3 star, accreting at very low accretion rates, from $10^{-12}\,\msunyr$ to $10^{-10}\,\msunyr$. The two spectral types are representative of the targets in our sample. As the line profiles also depend on the geometry and temperature, we also varied {\ri}, {\rw}, and {\tmax} within the ranges found in  Figure~\ref{fig:parameter_dist}. We fixed the inclination at 60{\degr}. In all cases, we assumed no contribution from the accretion shock. For each case, we include a contribution from the chromosphere with a Gaussian. We adopted the amplitude $a_c=1$ for K5, $a_c=3$ for M3, and $\sigma_c=25\,\kms$ for all spectral types. These values are consistent with the results in Table~\ref{tab:model_results_acc}.
 
\begin{figure*}[t]
\epsscale{1.0}
\plotone{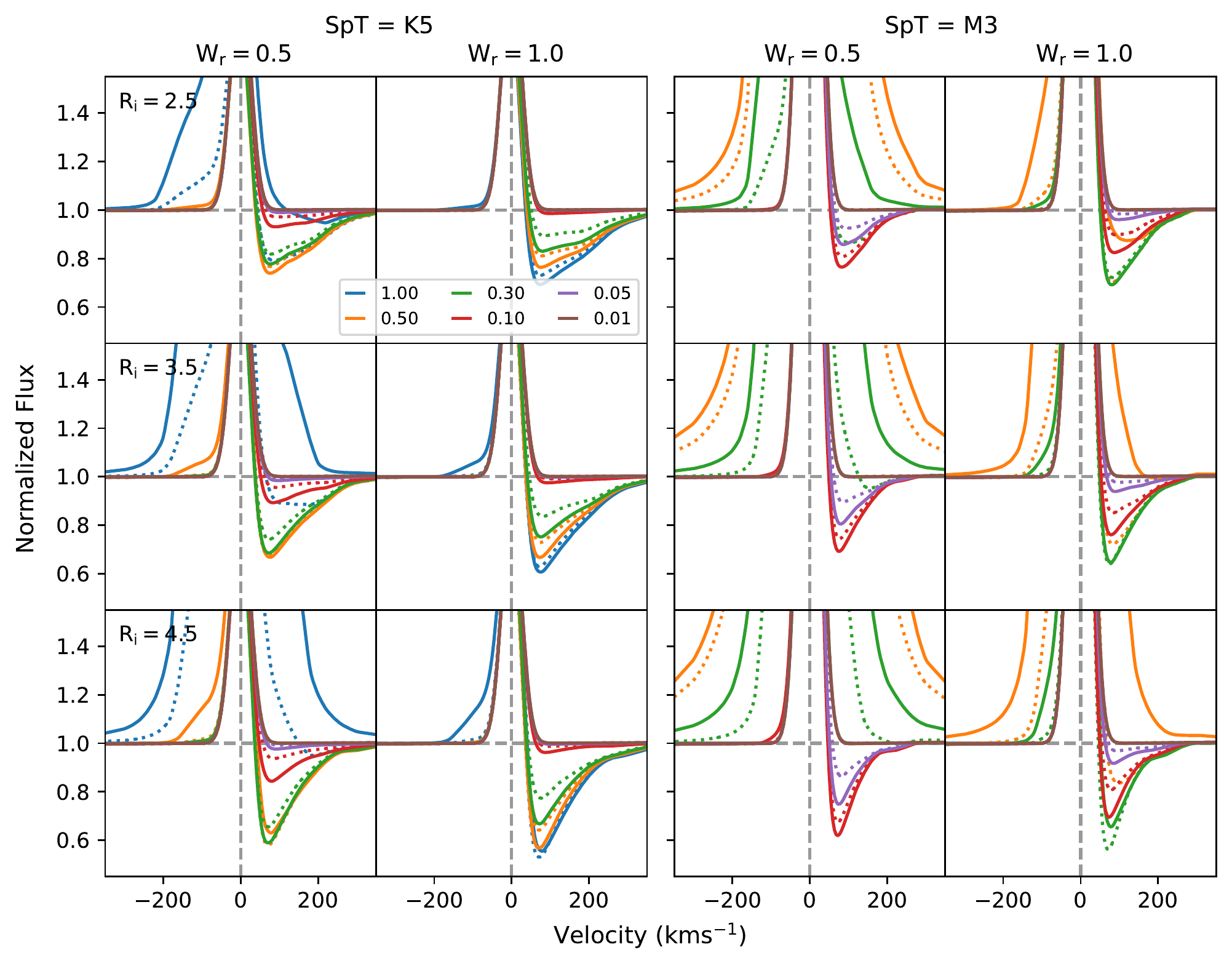}
\caption{The variation of line profiles at the lowest limit of accretion for stars with spectral types K5 and M3.
The two columns for each spectral type are for {\rw=0.5} and {\rw=1.0}, respectively. Each row corresponds to a fixed truncation radius {\ri}. Different colors are for different mass accretion rates; the legend shows the mass accretion rates in units of $10^{-10}\,\msunyr$. The solid lines are the profiles calculated for {\tmax}=13,000\,K and the dotted lines are for {\tmax}=12,000\,K. The black dash-dotted lines are the assumed chromospheric profiles. The range of the y-axis is fixed at low values to focus on the feature of the line near the continuum. The model at $\Mdot=10^{-12}\,\msunyr$, shown in brown, can barely be 
distinguishable from the chromosphere.
\label{fig:low_mdot}}
\end{figure*}
 
We show the results of the two small grids in Figure~\ref{fig:low_mdot}. In both stars, the emission component is stronger for large and narrow magnetospheres, as in \citet{muzerolle2001}. The redshifted absorption component of the line can be seen in all geometries for wide ranges of accretion rates. Assuming that we can detect a feature in the line that has $\sim30\%$ deviation from the continuum (i.e., signal-to-noise $\sim3$, see also CVSO~1928 in Fig.~\ref{fig:halpha_good}), we can estimate the lowest mass accretion rates detectable with each model by determining if the broad wings in emission is greater than $\sim1.3$ and the absorption feature is deeper than $\sim0.7$, both in the normalized unit of the continuum. For the K5 stars, we can see that the features would still be detectable at $\sim3-5\times10^{-11}\,\msunyr$ (green and orange lines) in all geometries considered. For the M3 stars, the lowest detectable mass accretion rate is $\sim1-3\times10^{-11}\,\msunyr$ (i.e., red and green lines).
 
The situation is more complicated in the real observations since chromospheric contributions could vary from object to object. However, if the accretion rates are truly low (but higher than $\sim 10^{-12}\,\msunyr$), the redshifted absorption will be very evident, as seen in all cases in Figure~\ref{fig:low_mdot}. These absorption features are much broader than the typical width of the chromospheric lines (Figure~\ref{fig:low_mdot}, Table~\ref{tab:model_results_acc}), and therefore accretion will still be detectable as long as the spectra are obtained at a sufficiently high resolution (we used R=32500, i.e., that of MIKE, in Figure~\ref{fig:low_mdot}).
 
The results from the two stars can also be approximately extrapolated; that is, the limit would be slightly lower than $3\times10^{-11}\,\msunyr$ for a spectral type later than M3, and slightly higher than $5\times10^{-11}\,\msunyr$ for a spectral type earlier than K5.
 
\subsection{How does accretion stop?} \label{ssec:acc_stop}
The analysis and discussion in previous sub-sections have provided some evidence of how accretion stops. In \S~\ref{ssec:acc_vs_star_prop}, we can see that essentially all low accretors are accreting in the unstable regime, but none is close to the propeller regime. Therefore, we can infer that accretion does not stop by the stellar rotation reaching the propeller regime, or at least that this process is not the main driver stopping accretion. Instead, other processes must be at play.
 
As discussed by \citet{thanathibodee2020} in the study of the PDS~70 system, the formation of giant planets may, in some way, contribute to decrease accretion onto the star. However, as shown in that work, the mass accretion rates of the two planets are very small compared to that of the host star, and therefore mass accretion onto planets is unlikely to be the main source stopping accretion. Nevertheless, the formation of giant planets creates large gaps in the protoplanetary disk; this could make disk dispersal processes such as photoevaporation become more pronounced \citep{alexander2014}.
 
The fact that most of the stars in our sample are accreting at $\sim10^{-10}\,\msunyr$ even though the model can essentially detect accretion an order of magnitude lower than that suggests that we have obtained a physical lower limit to the mass accretion rate in the mass range covered by the stars in our sample. Two possibilities may be able to explain this observational result.
 
First, it is possible that this is a selection bias, such that we only selected stars classified as accretors using the {\henir} line, and that the line is not sensitive below $10^{-10}\,\msunyr$. Therefore, we have missed stars accreting at $\sim10^{-11}\,\msunyr$. Continuous monitoring of possible accretors \citepalias[c.f.,][]{thanathibodee2022} could be used to explore this possibility.
 
One implication of the sensitivity problem is that we cannot rule out the possibility that the inner disk could host a gas reservoir with low viscosity \citep{hartmann2018}. In this case, the inner disk is first separated from the outer disk due to processes that deplete the outer disk (e.g., photoevaporation). Due to low viscosity, the inner disk will drain slowly and will last long. The mass accretion rate onto the star will decrease with time, and with a sufficiently sensitive probe, accretion could be detected at the level well below $10^{-10}\,\msunyr$. If $\sim10^{-10}\,\msunyr$ is the physical limit, the inner disk mass reservoir scenario can be reconciled with other processes removing gas from the disk, such as MHD winds launching from the inner disk \citep{hartmann2018,tabone2022b,tabone2022a}.

Second, it is also possible that $10^{-10}\,\msunyr$ is the actual limit to the rate at which a star can accrete mass from its inner disk. If the inner disk depletion timescale is short (due to, e.g., high viscosity), the mass accreted onto the star is then essentially controlled by the supply of mass from the outer disk. In this scenario, photoevaporation could play a crucial role, since photoevaporation models suggest that mass accretion onto the stars ceases in a very short time scale ($\sim10^5$\,yr) once the mass accretion rates equal the photoevaporative mass loss rates. With the steady supply of mass from the outer disk, controlled by the photoevaporative rate, our results suggest that the photoevaporative mass loss rates should not be higher than $1\times10^{-10}\,\msunyr$ for the mass range considered in this study ($0.1-1.2\,\msun$) at the last stages of disk evolution.
 
Models of photoevaporation as well as observational constraints suggest a wide range of photoevaporative mass loss rates from $\sim10^{-10}\,\msunyr$ to $10^{-8}\,\msunyr$, depending on which energy range dominates the disk irradiation (EUV, X-ray, or FUV; \citealt{clarke2001}, \citealt{owen2011}, \citealt{gorti2009a}). Our results agree with the low rates of mass accretion predicted by EUV-dominated photoevaporation, suggesting that it is likely the main mechanism depleting disks at their last stages of evolution. It is possible that the higher mass loss rates from other mechanism are needed at earlier stages of disk evolution, as the disk lifetime would be longer than observed if the EUV-dominated photoevaporation operates alone at all timescales \citep{alexander2014}.
 
Regardless of the mechanism removing mass from the disk, it is now quite clear that mass accretion onto the star stops due to the properties of the disk as opposed to those directly related to the stars, as previously suggested in \citetalias{thanathibodee2022}. It is possible that accretion stops because the disk simply runs out of gas, but the process of removing mass from the disk and essentially stopping accretion, remains unclear. Detailed studies of disk properties (e.g., density, mass, gap sizes, dust settling) and their related radiation fields (FUV, X-ray) are needed to pinpoint the most probable processes stopping accretion. This is a subject for a follow-up study.

\section{Summary and Conclusions} \label{sec:summary}
In this paper, we conducted a study of 22 accretors and two possible accretors identified as such using the profiles of {\henir} line. We applied the magnetospheric accretion model \citep{hartmann1994,muzerolle1998a,muzerolle2001} to fit the {\halpha} lines of the stars observed with moderate resolution spectrograph and inferred the accretion properties from the best fitting models. We summarize our findings as follows:
\begin{enumerate}
    \item Most of the {\halpha} lines have significant contributions from the narrow chromospheric emission component of the line, up to 60\% in some cases and $\sim30\%$ on average.  
    Representing the chromospheric emission with a Gaussian, we find that 20 stars can be fitted very well with the combined magnetospheric+chromospheric model. These stars are used for further analysis.
    \item A fraction of stars show low-velocity redshifted absorption on top of bright emission in the {\halpha} line profile. We speculate that the magnetospheres of these stars are similar to that of CVSO~1335 \citep{thanathibodee2019a}, where accretion occurs in a 
    complex, multi-flows geometry.
    \item We find an anticorrelation between
    the disk truncation radius and the mass accretion rates     in the M-type stars of the sample.    The power-law relation is consistent with theory and simulations within the uncertainty.
    \item The disk truncation radius and the mass accretion rates are not correlated in our small sample of K-type stars. If this relationship is confirmed by larger sample size, the result may suggest departures of the magnetic field from dipolar geometry in K-type low accretors.
    \item We calculated  corotation radii from the projected rotational velocity of the stars and found them to be always larger than the truncation radii, which implies  that the magnetic propeller  is not the main 
    agent stopping accretion.    We also find that essentially all low accretors accrete in the unstable regime. 
    \item We compared the magnetic radius calculated from mass accretion rates and stellar parameters with the truncation radius inferred from the model and found that either the efficiency parameter $\xi$ or the dipolar magnetic field B need to be 
    small for the two radii to be the same. For the efficiency parameters to be comparable to other studies, the dipole magnetic field must be much weaker than the total magnetic fields, assumed to be at a few kG. This suggests that, as stars evolve, most of the magnetic fields are in higher-order configurations, but the dipole component can still truncate the disk due to the low accretion rates.
    \item The inferred high inclinations of the majority of our targets suggest a possible observational bias introduced by adopting the equivalent width of {\halpha} as a selection criterion to identify low accretor candidates. This is because the equivalent width of the line depends strongly on the inclination for low accretors, such that it will often go below the threshold at high inclinations. Therefore, more low accretor candidates 
    get selected at high inclinations.
    \item Stars at low inclinations may also be susceptible to accretion misclassification since the chromospheric contribution to the emission is expected to be higher than in higher inclination stars. The chromospheric component may dominate subtle features (e.g., broad wings, weak absorption) of the magnetosphere.
    \item Mass accretion rates of low accretors are not much lower than $1\times10^{-10}\,\msunyr$, regardless of the mass of the star, even though we could detect rates 5 to 10 times lower. This suggests that photoevaporative mass loss rates 
   are of the same order, since otherwise these low accretion rates would not have been detected. This value for the photoevaporative rate is     consistent with the low end of EUV-driven photoevaporation. As we ruled out the propeller as the main driver stopping accretion, the same mass accretion rates across the mass range suggest that the processes that stop mass accretion onto the star occur in the disk. It is possible that accretion stops simply because the disk runs out of gas.
\end{enumerate}

\begin{acknowledgments}
\noindent
We thank the anonymous referee for the helpful comments and suggestions which improve the quality of the manuscript. We thank Lee Hartmann for his useful suggestions. We are grateful to Mario Mateo for coordinating the scheduling of the MIKE  observations, and to the staff at Las Campanas Observatory,  in face of the partial observatory closure due to the COVID-19 pandemic.  This project is supported in part by NASA grant NNX17AE57G. JH acknowledges support from program UNAM-DGAPA-PAPIIT grant IA102319. 
 
This paper is based on data obtained from the ESO Science Archive Facility. This research made use of Astropy,\footnote{http://www.astropy.org} a community-developed core Python package for Astronomy \citep{astropy2013,astropy2018}.
\end{acknowledgments}
 
\facilities{Magellan:Clay (MIKE), ESO:VLT (UVES, X-shooter)}

\software{CarPy, Astropy \citep{astropy2013,astropy2018}}

\bibliography{tts}{}
\bibliographystyle{aasjournal}

\end{document}